\documentclass{PoS}

\title{Planck and Electroweak Scales Emerging from Weyl Conformal Gravity}

\ShortTitle{Mass scales from conformal gravity}

\author{\speaker{Ichiro Oda}\\
           Department of Physics, Faculty of Science, University of the 
           Ryukyus,\\
           Nishihara, Okinawa 903-0213, Japan\\          
        E-mail: \email{ioda@sci.u-ryukyu.ac.jp}}

\abstract{We show that the Planck mass scale can be generated from conformal gravity in
the Weyl conformal geometry via the Coleman-Weinberg mechanism of dimensional transmutation
where quantum corrections stemming from the gravitational field and the Weyl gauge field trigger the 
symmetry breakdown of a local Weyl symmetry. It is also shown that the vacuum expectation value of a scalar field
is transmitted to a sector of the standard model through a potential involving the scale invariant
part and the contribution from the Coleman-Weinberg mechanism, thereby generating the electroweak 
scale.}

\FullConference{Corfu Summer Institute 2018 "School and Workshops on Elementary Particle Physics and Gravity"\\
		(CORFU2018)\\
		31 August - 28 September, 2018\\
		Corfu, Greece}

\begin{document}

\section{Introduction}

One of the most important problems in modern particle physics is to understand the origin of different mass scales
existing in nature. For instance, the electroweak mass scale $M_{EW}$ is generated via the Higgs mechanism where
the Higgs field takes a vacuum expectation value around $100 GeV$, while the QCD mass scale $\Lambda_{QCD}$,
which accounts for the masses of the nucleons and thus most of the visible mass in the universe (about $98 \%$ 
of the whole visible mass), is an example of the mass generation by dimensional transmutation.

In order to understand the origin of various mass scales, it is natural to start with a theory which has no intrinsic
mass scales and consider how the mass scales are generated from a massless world via a certain mechanism. 
It is well known that there is a local or global scale symmetry in such a theory without intrinsic mass scales. 
However, as stressed in \cite{Oda1}, if we couple the gravitational interaction to a theory, only local symmetries 
in general make sense since no-hair theorem \cite{MTW} of quantum black holes suggests that global additive 
conservation laws such as baryon and lepton number conservation cannot hold in any consistent
quantum gravity theory. Indeed, in string theory, we never get any additive conservation
laws and at least in known string vacua, the additive global symmetries turn out to be either
gauge symmetries or explicitly violated. By contrast, gauge symmetries such as U(1) electric
charge conservation law cause no trouble for black hole physics. Thus, in the gravitational
theory, the global symmetries should be promoted to the local gauge symmetries except 
some unbroken global symmetries associated with boundary or topology of four-dimensional manifolds.

In this article, we wish to construct a gravitational theory with the standard model (SM) where the local
scale symmetry plays a critical role, and then examine whether the Planck and the electroweak scales are
generated or not. To this aim, we will embed the SM and general relativity (GR) in a larger theory which
exhibits a local scale symmetry, and show that the Planck scale arises by the Coleman-Weinberg mechanism
via dimensional transmutation \cite{Coleman} and it is then transmitted to the SM through a scale invariant potential,
thereby generating the electroweak scale.   

A natural and general framework where a scale symmetry is implemented as a local gauge symmetry
is given by a generalization of the Riemann geometry, which is known as the Weyl geometry 
\cite{Weyl}.\footnote{See Ref. \cite{Scholz} for historical review on the Weyl geometry.}  The Weyl
geometry is defined as a geometry with a real symmetric metric tensor and a symmetric
connection given by Eq. (\ref{W-connection}) as seen shortly. It is therefore expected that 
when the Weyl gauge field $S_\mu$ is vanishing, the Weyl geometry becomes equivalent to the Riemann geometry. 
However, it turns out that if $S_\mu$ is a gradient, i.e., pure gauge, the Weyl geometry precisely reduces to 
the Riemann one. In the Weyl geometry, parallel displacement of a vector field changes its length so that 
the very notion of length becomes path-dependent. For instance, one can envisage a space traveller, who travels 
to a distant star and then returns to the earth, being surprised to know not only that people in the earth have aged 
much rather than him as predicted by GR in the Riemann geometry but also that the clock on the rocket runs at 
a different rate from that in the earth as understood by Weyl conformal gravity in the Weyl geometry, what we call, 
"the second clock problem" \cite{Penrose}. Based on this very striking geometry, Weyl has succeeded in geometrizing 
the electromagnetic theory in the space-time geometry.

Since we have already applied the idea of generating the Planck scale from the Coleman-Weinberg mechanism
to locally scale invariant gravitational theories with the SM in the Riemann geometry \cite{Oda1, Oda0}, in this article
we will deal with a locally scale invariant theory in the Weyl geometry and show that this is indeed the case even 
in this more general geometry. The detail will be reported in a separate publication in the near future.

\section{Weyl conformal geometry} 
 
We start with a brief review on the basic concepts and definitions of the Weyl conformal geometry.\footnote{See
also Refs. \cite{Fulton, Smolin, Cesare} for a concise introduction of the Weyl geometry.} We adopt the conventions 
and notation for the Riemann tensors and the metric signature in the Wald's textbook \cite{Wald}.   
 
 In the Weyl geometry, the Weyl transformation, which is the sum of the local conformal transformation for 
a generic field $\Phi (x)$ and the Weyl gauge transformation for a Weyl gauge field $S_\mu(x)$, is defined as
\begin{eqnarray}
\Phi (x) \rightarrow \Phi^\prime (x) = e^{w \Lambda(x)} \Phi (x), \qquad
S_\mu (x) \rightarrow S^\prime_\mu (x) = S_\mu (x) - \frac{1}{f} \partial_\mu \Lambda (x),
\label{Weyl transf}
\end{eqnarray}
where $w$ is the Weyl weight, $f$ is the coupling constant for the non-compact abelian gauge group, and 
$\Lambda(x)$ is the local parameter for the conformal transformation. Writing out the conformal transformation
for various fields explicitly,
\begin{eqnarray}
g_{\mu\nu} (x) &\rightarrow& g_{\mu\nu}^\prime (x) = e^{2 \Lambda(x)} g_{\mu\nu}(x), \qquad
\phi (x) \rightarrow \phi^\prime (x) = e^{- \Lambda(x)} \phi (x),  \nonumber\\
\psi (x) &\rightarrow& \psi^\prime (x) = e^{- \frac{3}{2} \Lambda(x)} \psi (x), \qquad
A_\mu (x) \rightarrow A^\prime_\mu (x) = A_\mu (x),
\label{Weyl transf 2}
\end{eqnarray}
where $g_{\mu\nu} (x)$, $\phi (x)$, $\psi (x)$ and $A_\mu (x)$ are the metric tensor, scalar, spinor,
and conventional gauge fields, respectively. Here it is convenient to define a Weyl covariant derivative $D_\mu$ for
a generic field $\Phi (x)$ with the Weyl weight $w$ as
\begin{eqnarray}
D_\mu \Phi \equiv \partial_\mu \Phi + w f S_\mu \Phi,
\label{W-cov-deriv}
\end{eqnarray}
which transforms covariantly under the Weyl transformation:
\begin{eqnarray}
D_\mu \Phi \rightarrow (D_\mu \Phi)^\prime = e^{w \Lambda(x)} D_\mu \Phi.
\label{S-cov-transf}
\end{eqnarray}

As mentioned above, the Weyl geometry is defined as a geometry with a real symmetric metric tensor $g_{\mu\nu}
(= g_{\nu\mu})$ and a symmetric connection $\tilde \Gamma^\lambda_{\mu\nu} (= \tilde \Gamma^\lambda_{\nu\mu})$ 
which is defined as
\begin{eqnarray}
\tilde \Gamma^\lambda_{\mu\nu} &=& \frac{1}{2} g^{\lambda\rho} \left( D_\mu g_{\nu\rho} + D_\nu g_{\mu\rho}
- D_\rho g_{\mu\nu} \right)
\nonumber\\
&=& \Gamma^\lambda_{\mu\nu} + f \left( S_\mu \delta^\lambda_\nu + S_\nu \delta^\lambda_\mu 
- S^\lambda g_{\mu\nu} \right),
\label{W-connection}
\end{eqnarray}
where 
\begin{eqnarray}
\Gamma^\lambda_{\mu\nu} \equiv \frac{1}{2} g^{\lambda\rho} \left( \partial_\mu g_{\nu\rho} 
+ \partial_\nu g_{\mu\rho} - \partial_\rho g_{\mu\nu} \right),
\label{Affine connection}
\end{eqnarray}
is the affine connection in the Riemann geometry. The most important difference between the Riemann geometry
and the Weyl one is that $\nabla_\lambda g_{\mu\nu} = 0$ (the metric condition) in the Riemann geometry, 
while in the Weyl geometry
\begin{eqnarray}
\tilde \nabla_\lambda g_{\mu\nu} \equiv \partial_\lambda g_{\mu\nu} - \tilde \Gamma^\rho_{\lambda\mu} 
g_{\rho\nu} - \tilde \Gamma^\rho_{\lambda\nu} g_{\mu\rho}
= - 2 f S_\lambda g_{\mu\nu}. 
\label{W-metric cond}
\end{eqnarray}
Let us recall that the metric condition implies that length and angle are preserved under parallel transport
whereas Eq. (\ref{W-metric cond}) does that only angle, but not length, is preserved by the Weyl connection.

Now using the Weyl connection $\tilde \Gamma^\lambda_{\mu\nu}$ one can construct a conformally 
invariant curvature tensor:
\begin{eqnarray}
\tilde R_{\mu\nu\rho} \, ^\sigma &\equiv& \partial_\nu \tilde \Gamma^\sigma_{\mu\rho} 
- \partial_\mu \tilde \Gamma^\sigma_{\nu\rho} + \tilde \Gamma^\alpha_{\mu\rho} \tilde \Gamma^\sigma_{\alpha\nu} 
- \tilde \Gamma^\alpha_{\nu\rho} \tilde \Gamma^\sigma_{\alpha\mu}
\nonumber\\
&=& R_{\mu\nu\rho} \, ^\sigma + f \left( \delta^\sigma_{[\mu} \nabla_{\nu]} S_\rho 
- \delta^\sigma_\rho \nabla_{[\mu} S_{\nu]} - g_{\rho [\mu} \nabla_{\nu]} S^\sigma \right)
\nonumber\\
&+& f^2 \left( S_{[\mu} \delta^\sigma_{\nu]} S_\rho - S_{[\mu} g_{\nu]\rho} S^\sigma
+ \delta^\sigma_{[\mu} g_{\nu]\rho} S_\alpha S^\alpha \right),
\label{W-curv-tensor}
\end{eqnarray}
where $R_{\mu\nu\rho} \, ^\sigma$ is the curvature tensor in the Riemann geometry, and
we have defined the antisymmetrization by the square bracket, e.g., $A_{[\mu} B_{\nu]} \equiv A_\mu B_\nu
- A_\nu B_\mu$. Then, it is straightforward to prove the following identities:
\begin{eqnarray}
\tilde R_{\mu\nu\rho} \, ^\sigma = - \tilde R_{\nu\mu\rho} \, ^\sigma,  \qquad
\tilde R_{[\mu\nu\rho]} \, ^\sigma = 0, \qquad
\tilde \nabla_{[\lambda} \tilde R_{\mu\nu]\rho} \, ^\sigma = 0.
\label{W-curv-identity}
\end{eqnarray}
The curvature tensor $\tilde R_{\mu\nu\rho} \, ^\sigma$ has $26$ independent components,
twenty of which are possessed by $R_{\mu\nu\rho} \, ^\sigma$ and six by the conformally
invariant field strength $H_{\mu\nu} \equiv \partial_\mu S_\nu - \partial_\nu S_\mu$.

From $\tilde R_{\mu\nu\rho} \, ^\sigma$ it is possible to define a conformally invariant 
Ricci tensor:
\begin{eqnarray}
\tilde R_{\mu\nu} &\equiv& \tilde R_{\mu\rho\nu} \, ^\rho
\nonumber\\
&=& R_{\mu\nu} + f \left( - 2 \nabla_\mu S_\nu - H_{\mu\nu} - g_{\mu\nu} \nabla_{\alpha} S^\alpha \right)
+ 2 f^2 \left( S_\mu S_\nu - g_{\mu\nu} S_\alpha S^\alpha \right).
\label{W-Ricci-tensor}
\end{eqnarray}
Let us note that 
\begin{eqnarray}
\tilde R_{[\mu\nu]} \equiv \tilde R_{\mu\nu} - \tilde R_{\nu\mu} = - 4 f H_{\mu\nu}.
\label{W-Ricci-tensor 2}
\end{eqnarray}
Similarly, one can define a conformally not invariant but covariant scalar curvature:
\begin{eqnarray}
\tilde R \equiv g^{\mu\nu} \tilde R_{\mu\nu} 
= R - 6 f \nabla_\mu S^\mu - 6 f^2 S_\mu S^\mu.
\label{W-scalar-curv}
\end{eqnarray}
We find that under the Weyl transformation  (\ref{Weyl transf}), $\tilde R \rightarrow \tilde R^\prime = e^{- 2 \Lambda(x)}
\tilde R$ while $\tilde \Gamma^\lambda_{\mu\nu}, \tilde R_{\mu\nu\rho} \, ^\sigma$ and $\tilde R_{\mu\nu}$
are all invariant.

Finally, we wish to write out a generalization of the Gauss-Bonnet topological invariant in the Weyl geometry 
which can be described as \cite{Drechsler}
\begin{eqnarray}
I_{GB} &\equiv& \int d^4 x \sqrt{-g} \, \epsilon^{\mu\nu\rho\sigma} \epsilon_{\alpha\beta\gamma\delta} \, 
\tilde R_{\mu\nu} \, ^{\alpha\beta} \tilde R_{\rho\sigma} \, ^{\gamma\delta}    
\nonumber\\
&=& - 2 \int d^4 x \sqrt{-g} \,  \left( \tilde R_{\mu\nu\rho\sigma} \tilde R^{\rho\sigma\mu\nu} 
- 4 \tilde R_{\mu\nu} \tilde R^{\nu\mu} + \tilde R^2 - 12 f^2 H_{\mu\nu} H^{\mu\nu} \right)
\nonumber\\
&=& - 2 \int d^4 x \sqrt{-g} \,  \left( R_{\mu\nu\rho\sigma} R^{\mu\nu\rho\sigma} 
- 4 R_{\mu\nu} R^{\mu\nu} + R^2 \right).   
\label{GB}
\end{eqnarray}

\section{Weyl invariant quadratic gravity} 

In this section, we would like to look for a gravitational theory with the SM based on the Weyl geometry 
outlined in the previous section. It is of interest to notice that if only the metric tensor is allowed to use for 
the construction of  a gravitational action, the action invariant under the Weyl transformation is restricted to be 
of form of quadratic gravity, but not be of the Einstein-Hilbert type. Using the topological invariant (\ref{GB}), 
up to the kinetic term for the Weyl gauge field, one can write out a general action of quadratic gravity, 
which is invariant under the Weyl transformation, as follows:
\begin{eqnarray}
S_{QG} = \int d^4 x \sqrt{-g} \left[ - \frac{1}{2 \xi^2}  \tilde C_{\mu\nu\rho\sigma} \tilde C^{\mu\nu\rho\sigma} 
+ \frac{\lambda}{4 !} \tilde R^2 \right] \equiv \int d^4 x \sqrt{-g} \, {\cal{L}}_{QG},
\label{QG}
\end{eqnarray}
where $\xi$ and $\lambda$ are coupling constants. Moreover, a generalization of the conformal tensor, 
$\tilde C_{\mu\nu\rho\sigma}$, in the Weyl geometry is defined as in $C_{\mu\nu\rho\sigma}$ in the Riemann geometry:
\begin{eqnarray}
\tilde C_{\mu\nu\rho\sigma} &\equiv& \tilde R_{\mu\nu\rho\sigma} - \frac{1}{2} \left( g_{\mu\rho} \tilde R_{\nu\sigma}
+ g_{\nu\sigma} \tilde R_{\mu\rho} -  g_{\mu\sigma} \tilde R_{\nu\rho} - g_{\nu\rho} \tilde R_{\mu\sigma} \right)
+ \frac{1}{6} \left( g_{\mu\rho} g_{\nu\sigma} - g_{\mu\sigma} g_{\nu\rho} \right) \tilde R
\nonumber\\
&=& C_{\mu\nu\rho\sigma} + f \left[ - g_{\rho\sigma} H_{\mu\nu} + \frac{1}{2} \left( g_{\mu\rho} H_{\nu\sigma}
+  g_{\nu\sigma} H_{\mu\rho} - g_{\mu\sigma} H_{\nu\rho} - g_{\nu\rho} H_{\mu\sigma} \right) \right].
\label{Conformal tensor}
\end{eqnarray}
This conformal tensor in the Weyl geometry has the following properties:
\begin{eqnarray}
\tilde C_{\mu\nu\rho\sigma} = - \tilde C_{\nu\mu\rho\sigma}, \qquad 
\tilde C_{\mu\nu\rho} \, ^\nu = 0, \qquad 
\tilde C_{\mu\nu\rho} \, ^\rho = - 4 f H_{\mu\nu}.
\label{Conformal tensor 2}
\end{eqnarray}

Next, by introducing a scalar field $\phi$ and using the classical equivalence, let us rewrite $\tilde R^2$ in the action 
(\ref{QG}) in the form of the scalar-tensor gravity plus $\lambda \phi^4$ interaction \cite{Ghilencea} whose
Lagrangian density takes the form
\begin{eqnarray}
\frac{1}{\sqrt{-g}} {\cal{L}}_{QG} &=& - \frac{1}{2 \xi^2}  \tilde C_{\mu\nu\rho\sigma} \tilde C^{\mu\nu\rho\sigma} 
+ \frac{\lambda}{12} \phi^2 \tilde R - \frac{\lambda}{4 !} \phi^4 
\nonumber\\
&=& - \frac{1}{2 \xi^2}  \tilde C_{\mu\nu\rho\sigma} \tilde C^{\mu\nu\rho\sigma} 
+ \frac{1}{12} \phi^2 \tilde R - \frac{\lambda_\phi}{4 !} \phi^4 
\nonumber\\
&=& - \frac{1}{2 \xi^2}  C_{\mu\nu\rho\sigma} C^{\mu\nu\rho\sigma} 
+ \frac{1}{12} \phi^2 R - \frac{\lambda_\phi}{4 !} \phi^4 - \frac{3 f^2}{\xi^2} H_{\mu\nu}^2 
\nonumber\\
&-& \frac{1}{2} \phi^2 ( f \nabla_\mu S^\mu + f^2 S_\mu S^\mu ),  
\label{QG 2}
\end{eqnarray}
where in the second equality we have redefined $\sqrt{\lambda} \phi \rightarrow \phi$ and set 
$\lambda = \frac{1}{\lambda_\phi}$. Here it is straightforward to write down a standard model (SM) or physics
beyond the standard model (BSM) action which is invariant under the Weyl transformation, but we will omit to 
tough on it in this article and present the detail in a separate publication.\footnote{Some related models on the
basis of the Weyl geometry have been made in Refs. \cite{Cheng, Nishino1, Nishino2}.}

\section{Emergence of Planck scale}

At low energies, general relativity (GR) describes a lot of gravitational and astrophysical phenomena neatly,
so the Weyl invariant Lagrangian density (\ref{QG 2}) of quadratic gravity should be reduced to that of GR at low
energies. To do that, we need to break the Weyl symmetry at any rate by some method. The standard procedure
done so far is to take a gauge condition for the Weyl transformation such that $\phi = \phi_0$ where $\phi_0$ 
is a certain constant \cite{Smolin, Cesare, Ghilencea}. However, $\phi_0$ is a free parameter which is not fixed 
by theory in itself so it is not clear why we choose a specific value $\phi_0 \sim M_{Pl}$ where $M_{Pl}$ is the Planck mass 
scale defined as $M_{Pl} = \frac{1}{\sqrt{8 \pi G}} = 2.44 \times 10^{18} GeV$ with $G$ being the Newton constant.     
We wish to construct a theory where the scalar field $\phi$ acquires a vacuum expectation value (VEV) as a result 
of instabilities in the full quantum theory including quantum corrections from gravity. Technically speaking, what
we expect is that after quantum corrections are included the effective potential has a form favoring the specific
VEV, $\langle \phi \rangle \sim M_{Pl}$.

To this aim, let us first expand the scalar field and the metric around a classical field $\phi_c$ and a flat Minkowski
metric $\eta_{\mu\nu}$ like \cite{Oda1}
\begin{eqnarray}
\phi = \phi_c + \varphi,    \quad 
g_{\mu\nu} = \eta_{\mu\nu} + \xi h_{\mu\nu},
\label{Expansion}
\end{eqnarray}
where we take $\phi_c$ to be a constant since we are interested in the effective potential depending on
the constant $\phi_c$. Next, since we wish to calculate the one-loop effective potential, we will
derive only quadratic terms in quantum fields from the classical Lagrangian density (\ref{QG 2}).
Then, up to surface terms the Lagrangian density corresponding to the conformal tensor squared takes the form
\begin{eqnarray}
{\cal L}_C \equiv - \frac{1}{2 \xi^2}  \sqrt{-g} \, C_{\mu\nu\rho\sigma} C^{\mu\nu\rho\sigma} 
= - \frac{1}{4} h^{\mu\nu} P^{(2)}_{\mu\nu, \rho\sigma} \Box^2 h^{\rho\sigma},
\label{Weyl Lagr}
\end{eqnarray}
where $P^{(2)}_{\mu\nu, \rho\sigma}$ is the projection operator for spin-2 modes\footnote{We follow
the definition of projection operators in \cite{Nakasone1, Nakasone2}.} and $\Box \equiv \eta^{\mu\nu}
\partial_\mu \partial_\nu$. In a similar manner, the Lagrangian density corresponding to the scalar-tensor
gravity in Eq. (\ref{QG 2}) reads
\begin{eqnarray}
{\cal L}_{ST} &\equiv& \sqrt{-g} \, \frac{1}{12} \phi^2 R
\nonumber\\
&=& \frac{1}{48} \xi^2 \phi^2_c h^{\mu\nu} \left( P^{(2)}_{\mu\nu, \rho\sigma} 
- 2 P^{(0, s)}_{\mu\nu, \rho\sigma} \right) \Box h^{\rho\sigma} 
- \frac{1}{6} \xi \phi_c \varphi \left( \eta_{\mu\nu} - \frac{1}{\Box} \partial_\mu \partial_\nu \right) 
\Box h^{\mu\nu}.
\label{ST-Lagr}
\end{eqnarray}
The remaining Lagrangian density can be evaluated in a similar way and consequently all the quadratic terms 
in (\ref{QG 2}) are summarized to be
\begin{eqnarray}
{\cal L}_{QG} &=& \frac{1}{4} h^{\mu\nu} \left[ \left( - \Box + \frac{1}{12} \xi^2 \phi^2_c \right)
P^{(2)}_{\mu\nu, \rho\sigma} - \frac{1}{6} \xi^2 \phi^2_c P^{(0, s)}_{\mu\nu, \rho\sigma} \right]
\Box h^{\rho\sigma} 
\nonumber\\
&-& \frac{1}{6} \xi \phi_c \varphi \left( \eta_{\mu\nu} 
- \frac{1}{\Box} \partial_\mu \partial_\nu \right) \Box h^{\mu\nu}
- \frac{\lambda_\phi}{4} \phi_c^2 \varphi^2 -  \frac{\lambda_\phi}{12} \xi \phi_c^3 h \varphi
\nonumber\\
&-& \frac{3 f^2}{\xi^2} H^{\prime 2}_{\mu\nu} - \frac{1}{2} f^2 \phi_c^2 S^\prime_\mu S^{\prime\mu}
- \frac{1}{2} \varphi \Box \varphi,
\label{Total-Lagr}
\end{eqnarray}
where we have set $S^\prime_\mu = S_\mu - \frac{1}{f \phi_c} \partial_\mu \varphi$ and 
$H^\prime_{\mu\nu} = \partial_\mu S^\prime_\nu - \partial_\nu S^\prime_\mu$.

At this stage, let us set up the gauge-fixing conditions. For diffeomorphisms, we adopt a gauge condition
\begin{eqnarray}
\chi_\mu \equiv \xi \partial^\nu ( h_{\mu\nu} - \frac{1}{4} \eta_{\mu\nu} h ) = 0,
\label{Gauge-diffo}
\end{eqnarray}
which is invariant under the conformal transformation. We find that the Lagrangian density for this gauge condition
and its corresponding FP ghost term is given by 
\begin{eqnarray}
{\cal L}_{GF + FP} = - \frac{1}{2 \alpha} \chi_\mu Y^{\mu\nu} \chi_\nu 
- \bar c_\mu Y^{\mu\nu} ( \Box c_\nu + \frac{1}{2} \partial_\nu \partial_\rho c^\rho),
\label{Gauge+FP}
\end{eqnarray}
where $\alpha$ is a gauge parameter. Let us recall that in case of the higher derivative gravity 
\cite{Stelle, Tonin}, it is convenient to work with a more general gauge-fixing term ${\cal L}_{GF} 
= - \frac{1}{2 \alpha} \chi_\mu Y^{\mu\nu} \chi_\nu$ where $Y^{\mu\nu}$ is the weight function 
involving derivatives and gauge parameters \cite{Fradkin, Odintsov}.  For instance, a suitable choice is
$Y^{\mu\nu} = \eta^{\mu\nu} \Box + c \partial^\mu \partial^\nu$ ($c$ is a constant) and in calculating
the propagators, the constants $c$ and $\alpha$ are fixed to be an appropriate value for simplifying 
the expressions. Next, we fix the gauge freedom corresponding to the Weyl transformation by a gauge condition
\begin{eqnarray}
h \equiv \eta^{\mu\nu} h_{\mu\nu}  = 0.
\label{Gauge-conf}
\end{eqnarray}
Since this gauge fixing condition and its Weyl transformation contain no derivatives, one can 
neglect the corresponding FP ghost term in the one-loop approximation. 

Consequently, we can obtain a quantum Lagrangian density involving only quadratic terms:
\begin{eqnarray}
{\cal L}_{QG} &=& \frac{1}{4} h^{\mu\nu} \left( - \Box + \frac{1}{12} \xi^2 \phi^2_c \right) 
P^{(2)}_{\mu\nu, \rho\sigma} \Box h^{\rho\sigma} 
- \frac{1}{72} \xi^2 \phi^2_c \partial_\mu \partial_\nu h^{\mu\nu} \frac{1}{\Box} 
\partial_\rho \partial_\sigma h^{\rho\sigma}
\nonumber\\
&+& \frac{1}{6} \xi \phi_c \varphi \partial_\mu \partial_\nu h^{\mu\nu}
- \frac{1}{2} \varphi \left( \Box + \frac{1}{2} \lambda_\phi \phi^2_c \right) \varphi
- \frac{1}{4} \hat H_{\mu\nu}^2 - \frac{1}{2} \frac{\xi^2 \phi^2_c}{12} \hat S_\mu \hat S^\mu
\nonumber\\
&-& \frac{1}{2 \alpha} \chi_\mu Y^{\mu\nu} \chi_\nu 
- \bar c_\mu Y^{\mu\nu} ( \Box c_\nu + \frac{1}{2} \partial_\nu \partial_\rho c^\rho),
\label{Q-Lagr}
\end{eqnarray}
where we have introduced $\hat S_\mu \equiv \frac{2 \sqrt{3} f}{\xi} S^\prime_\mu$,
defined $\hat H_{\mu\nu} \equiv \partial_\mu \hat S_\nu - \partial_\nu \hat S_\mu$ and used
the gauge condition  (\ref{Gauge-conf}). Moreover, in order to eliminate the mixed term involving $\varphi$ 
and $h_{\mu\nu}$, let us introduce a new field $\hat \varphi$ instead of $\varphi$, which is defined as
\begin{eqnarray}
\hat \varphi \equiv \varphi - \frac{1}{6} \xi \phi_c \left( \Box 
+ \frac{1}{2} \lambda_\phi \phi^2_c \right)^{-1} \partial_\mu \partial_\nu h^{\mu\nu}.
\label{varphi-hat}
\end{eqnarray}
Then, the Lagrangian density (\ref{Q-Lagr}) can be rewritten as
\begin{eqnarray}
{\cal L}_{QG} &=& \frac{1}{4} h^{\mu\nu} \left( - \Box + \frac{1}{12} \xi^2 \phi^2_c \right) 
P^{(2)}_{\mu\nu, \rho\sigma} \Box h^{\rho\sigma} 
- \frac{1}{4} \hat H_{\mu\nu}^2 - \frac{1}{2} \frac{\xi^2 \phi^2_c}{12} \hat S_\mu \hat S^\mu
\nonumber\\
&-& \frac{1}{2} \hat \varphi \left( \Box + \frac{1}{2} \lambda_\phi \phi^2_c \right) \hat \varphi
- \frac{1}{144} \lambda_\phi \xi^2 \phi^4_c \partial_\mu \partial_\nu h^{\mu\nu} 
\frac{1}{\Box \left(\Box + \frac{1}{2} \lambda_\phi \phi^2_c \right)} 
\partial_\rho \partial_\sigma h^{\rho\sigma}
\nonumber\\
&-& \frac{1}{2 \alpha} \chi_\mu Y^{\mu\nu} \chi_\nu 
- \bar c_\mu Y^{\mu\nu} ( \Box c_\nu + \frac{1}{2} \partial_\nu \partial_\rho c^\rho).
\label{Q-Lagr2}
\end{eqnarray}

Based on this quantum Lagrangian density (\ref{Q-Lagr2}), we can evaluate the one-loop effective action
by integrating out quantum fluctuations.  Then, up to a classical potential, the effective action $\Gamma [\phi_c]$ reads
\begin{eqnarray}
\Gamma [\phi_c] = i \frac{5 + 3}{2} \log \mathrm{det} \left( - \Box + \frac{1}{12} \xi^2 \phi^2_c \right).
\label{EA}
\end{eqnarray}
Here some remarks are in order. First, in this expression, the factors $5$ and $3$ come from the fact that 
a massive spin-2 state and a massive spin-1 Weyl gauge field possess five and three physical degrees of freedom, 
respectively. In order to see the pole structure of the massive Weyl gauge field $\hat S_\mu$ more easily,
it might be useful to introduce a scalar field $\pi$ to ensure the gauge symmetry. Indeed, redefining 
$\hat S_\mu \rightarrow \hat S_\mu + \partial_\mu \pi$ and then integrating over $\pi$ leads to the following 
gauge invariant expression:     
\begin{eqnarray}
{\cal L}_S &\equiv&  - \frac{1}{4} \hat H_{\mu\nu}^2 - \frac{1}{2} \frac{\xi^2 \phi^2_c}{12} (\hat S_\mu 
+ \partial_\mu \pi )^2
\nonumber\\
&\rightarrow& \frac{1}{4} \hat H_{\mu\nu} \frac{ -\Box + \frac{\xi^2 \phi^2_c}{12}}{\Box} \hat H^{\mu\nu}.
\label{Pi-field}
\end{eqnarray}
Taking the Lorenz gauge condition $\partial_\mu \hat S^\mu = 0$, the Lagrangian density becomes
\begin{eqnarray}
{\cal L}_S = - \frac{1}{2} \hat S_\mu (-\Box + \frac{\xi^2 \phi^2_c}{12}) \hat S^\mu.
\label{Pi-field2}
\end{eqnarray}
This form clearly shows that the massive gauge field at hand has the mass squared $\frac{\xi^2 \phi^2_c}{12}$,
but one disadvantage is that not three but two degrees of freedom has such a property. The remaining one
degree of freedom is contained in the longitudinal mode which is integrated over in the above derivation. 
Finally, it is valuable to note that the determinant $\mathrm{det} \left( \Box + \frac{1}{2} \lambda_\phi \phi^2_c \right)$ 
coming from the scalar field $\hat \varphi$ is precisely cancelled by that from the metric tensor, which implies that
"dilaton" $\hat \varphi$ is no longer a dynamical mode and is absorbed into the logitudinal mode of the Weyl gauge field,
thereby the massless gauge field becoming a massive one with mass squared equal to $\frac{\xi^2 \phi^2_c}{12}$.   
Also note that we have ignored the part of the effective action which is independent of $\phi_c$ since it never
gives us the effective potential for $\phi_c$.

To calculate $\Gamma [\phi_c]$, we will proceed step by step: First, let us note that $\Gamma [\phi_c]$ can be 
rewritten as follows:
\begin{eqnarray}
\Gamma [\phi_c] &=& 4 i \, \mathrm{Tr} \log \left( - \Box + \frac{1}{12} \xi^2 \phi^2_c \right)
\nonumber\\
&=& 4 i \int d^4 x \, \langle x|  \log \left( - \Box + \frac{1}{12} \xi^2 \phi^2_c \right) | x \rangle
\nonumber\\
&=& 4 i \int d^4 x \int \frac{d^4 k}{(2 \pi)^4} \, \langle x|  \log \left( - \Box + \frac{1}{12} \xi^2 \phi^2_c \right) 
| k \rangle \langle k | x \rangle
\nonumber\\
&=& 4 i (VT) \int \frac{d^4 k}{(2 \pi)^4} \log \left( k^2 + \frac{1}{12} \xi^2 \phi^2_c \right)
\nonumber\\
&=& 4 (VT) \frac{\Gamma(- \frac{d}{2})}{(4 \pi)^{\frac{d}{2}}} \left( \frac{1}{12} \xi^2 \phi^2_c \right)^{\frac{d}{2}},
\label{EA2}
\end{eqnarray}
where $(VT)$ denotes the space-time volume and in the last equality we have used the Wick rotation and 
the dimensional regularization.

Next, let us evaluate $\Gamma [\phi_c]$ in terms of the modified minimal subtraction scheme. In this scheme, 
the $\frac{1}{\varepsilon}$ poles (where $\varepsilon \equiv 4 - d$) together with the Euler-Mascheroni constant
$\gamma$ and $\log (4 \pi)$ are subtracted and then replaced with $\log M^2$ where $M$ is an arbitrary
mass parameter which is introduced to make the final equation dimensionally correct \cite{Peskin}. 
By subtracting the $\frac{1}{\varepsilon}$ pole, (\ref{EA2}) is reduced to the form
\begin{eqnarray}
- \frac{1}{VT} \Gamma [\phi_c] &=& - 4 \frac{\Gamma( 2 - \frac{d}{2} )}{\frac{d}{2} (\frac{d}{2} - 1)}
\frac{1}{(4 \pi)^{\frac{d}{2}}} \left( \frac{1}{12} \xi^2 \phi^2_c \right)^{\frac{d}{2}}
\nonumber\\
&=& - \frac{4}{2 (4 \pi)^2} \left( \frac{1}{12} \xi^2 \phi^2_c \right)^2 \left[ \frac{2}{\varepsilon} - \gamma
+ \log (4 \pi)  - \log \left(\frac{1}{12} \xi^2 \phi^2_c \right) + \frac{3}{2} \right]
\nonumber\\
&\rightarrow& \frac{2}{(4 \pi)^2} \left( \frac{1}{12} \xi^2 \phi^2_c \right)^2 
\left[ \log \left(\frac{\xi^2 \phi^2_c}{12 M^2} \right) - \frac{3}{2} \right].
\label{EA3}
\end{eqnarray}
Then, the one-loop effective potential will be of form\footnote{At first sight, the existence of the
$c_2 \phi^2_c$ might appear to be strange, but this term in fact emerges in the cutoff regularization. Note that
the only logarithmically divergent term, but not quadratic divergent one, arises in the dimensional regularization.}
\begin{eqnarray}
V_{eff}^{(1)} (\phi_c) = c_1 + c_2 \phi^2_c + \frac{1}{1152 \pi^2} \xi^4 \phi^4_c \log \left(\frac{\phi^2_c}{c_3}\right),
\label{EP}
\end{eqnarray}
where $c_i (i = 1, 2, 3)$ are constants to be determined by the renormalization conditions:
\begin{eqnarray}
\left. V_{eff}^{(1)} \right\vert_{\phi_c = 0} = \left. \frac{d^2 V_{eff}^{(1)}}{d \phi^2_c} \right\vert_{\phi_c = 0} 
= \left. \frac{d^4 V_{eff}^{(1)}}{d \phi^4_c} \right\vert_{\phi_c = \mu} = 0,
\label{Ren-cond}
\end{eqnarray}
where $\mu$ is the renormalization mass. As a result, we have the one-loop effective potential
\begin{eqnarray}
V_{eff}^{(1)} (\phi_c) = \frac{1}{1152 \pi^2} \xi^4 \phi^4_c \left( \log \frac{\phi^2_c}{\mu^2} - \frac{25}{6} \right).
\label{EP2}
\end{eqnarray}
Finally, by adding the classical potential we can arrive at the effective potential in the one-loop approximation
\begin{eqnarray}
V_{eff} (\phi_c) = \frac{\lambda_\phi}{4 !} \phi^4_c + \frac{1}{1152 \pi^2} \xi^4 \phi^4_c 
\left( \log \frac{\phi^2_c}{\mu^2} - \frac{25}{6} \right).
\label{EP3}
\end{eqnarray}

It is easy to see that this effective potential has a minimum at $\phi_c = \langle \phi \rangle$ away from the origin 
where the effective potential, $V_{eff} (\langle \phi \rangle)$, is negative. Since the renormalization mass $\mu$ 
is arbitrary, we will choose it to be the actual location of the minimum, $\mu = \langle \phi \rangle$ \cite{Coleman}:
\begin{eqnarray}
V_{eff} (\phi_c) = \frac{\lambda_\phi}{4 !} \phi^4_c + \frac{1}{1152 \pi^2} \xi^4 \phi^4_c 
\left( \log \frac{\phi^2_c}{\langle \phi \rangle^2} - \frac{25}{6} \right).
\label{EP4}
\end{eqnarray}
Since $\phi_c = \langle \phi \rangle$ is defined to be the minimum of $V_{eff}$, we deduce
\begin{eqnarray}
0 &=& \left. \frac{d V_{eff}}{d \phi_c} \right\vert_{\phi_c = \langle \phi \rangle}
\nonumber\\ 
&=& \left( \frac{\lambda_\phi}{6} - \frac{11}{864 \pi^2} \xi^4 \right) \langle \phi \rangle^3,
\label{Min-cond}
\end{eqnarray}
or equivalently, 
\begin{eqnarray}
\lambda_\phi = \frac{11}{144 \pi^2} \xi^4.
\label{Min-cond2}
\end{eqnarray}
This relation is similar to $\lambda = \frac{33}{8 \pi^2} e^4$ in case of the scalar QED in 
Ref. \cite{Coleman}, so as in that paper, the perturbation theory holds for very small $\xi$ as well.

The substitution of Eq. (\ref{Min-cond2}) into $V_{eff}$ in (\ref{EP4}) leads to
\begin{eqnarray}
V_{eff} (\phi_c) = \frac{1}{1152 \pi^2} \xi^4 \phi^4_c 
\left( \log \frac{\phi^2_c}{\langle \phi \rangle^2} - \frac{1}{2} \right).
\label{EP5}
\end{eqnarray}
Thus, the effective potential is now parametrized in terms of $\xi$ and $\langle \phi \rangle$
instead of $\xi$ and $\lambda_\phi$; it is nothing but $\it{dimensional \, transmutation}$, i.e., a dimensionless
coupling constant $\lambda_\phi$ is traded for a dimensional quantity $\langle \phi \rangle$ via
symmetry breakdown of the $\it{local}$ Weyl symmetry.   

Hence, from the classical Lagrangian density (\ref{QG 2}) of quadratic gravity, via dimensional transmutation,
the Einstein-Hilbert term for GR is induced in such a way that the Planck mass $M_{Pl}$ is given by
\begin{eqnarray}
M_{Pl}^2 = \frac{1}{6} \langle \phi \rangle^2.
\label{Planck mass}
\end{eqnarray}
At the same time, the Weyl gauge field becomes massive by 'eating' the dilaton $\varphi$ whose magnitude
of mass is given 
\begin{eqnarray}
m_S^2 = \frac{1}{12} \xi^2 \langle \phi \rangle^2 = \frac{1}{2} \xi^2 M_{Pl}^2.
\label{Gauge mass}
\end{eqnarray}
As long as the perturbation theory is concerned, the coupling constant $\xi$ must take a small value,
$\xi \ll 1$. At the low energy region satisfying $E \ll m_S$, we can integrate over the massive Weyl gauge field,
and consequently not only we would have GR with the SM but also the second clock effect has no physical
effects at low energies.

\section{Emergence of electroweak scales}

Though we have ignored the existence of the SM and limited ourselves to the gravitational sector thus far, 
it is straightforward to incorporate the SM or the BSM in the present formulation by replacing the gauge covariant derivative 
with the Weyl covariant one. As for the Higgs sector, since the (tachyonic) Higgs mass term breaks the scale symmetry, 
after the symmetry breakdown of the local Weyl symmetry by the Coleman-Weinberg mechanism, the Higgs potential 
must be replaced by a potential 
\begin{eqnarray}
V_{eff} (\phi, H) = \frac{1}{1152 \pi^2} \xi^4 \phi^4 \left( \log \frac{\phi^2}{\langle \phi \rangle^2} 
- \frac{1}{2} \right) + \lambda_{H \phi} (H^\dagger H) \phi^2  
+ \frac{\lambda_H}{2} (H^\dagger H)^2.
\label{H-EP}
\end{eqnarray}
Inserting the minimum $\phi = \langle \phi \rangle$ to Eq. (\ref{H-EP}) and completing the square, 
the effective potential reduces to
\begin{eqnarray}
V_{eff} (\langle \phi \rangle, H) = \frac{\lambda_H}{2} \left( H^\dagger H 
+ \frac{\lambda_{H \phi}}{\lambda_H} \langle \phi \rangle^2 \right)^2
- \frac{1}{2} \left( \frac{\lambda^2_{H \phi}}{\lambda_H} + \frac{1}{1152 \pi^2} \xi^4 \right)
\langle \phi \rangle^4.
\label{H-EP2}
\end{eqnarray}
Owing to $\lambda_H > 0$, this potential has a minimum at $H^\dagger H = -\frac{\lambda_{H \phi}}{\lambda_H} 
\langle \phi \rangle^2$. Taking the unitary gauge $H^T = \frac{1}{\sqrt{2}} (0, v + h)$, this fact implies that
the squares of the VEV $v$ and the Higgs mass $m_h$ are given by
\begin{eqnarray}
v^2 = \frac{2 | \lambda_{H \phi} |}{\lambda_H} \langle \phi \rangle^2,  \quad 
m_h^2 = \lambda_H v^2.
\label{v and H-mass}
\end{eqnarray}
Using Eqs.  (\ref{Planck mass}) and  (\ref{v and H-mass}), the magnitude of the coupling constant $\lambda_{H \phi}$
reads
\begin{eqnarray}
| \lambda_{H \phi} |  = \frac{1}{12} \left( \frac{m_h}{M_{pl}} \right)^2 \sim {\cal O} (10^{-33}).
\label{lambda}
\end{eqnarray}
This relation $| \lambda_{H \phi} | \ll 1$ implies that the Higgs sector (or the SM sector) almost decouples from the 
gravitational sector but the mixed term $\lambda_{H \phi} (H^\dagger H) \phi^2$ in the potential (\ref{H-EP})
plays an important role in the sense that the VEV,  $\langle \phi \rangle$, obtained by the Coleman-Weinberg mechanism 
around the Planck mass scale, is transmitted to the SM sector, thereby generating the Higgs mass term.

\section{Conclusions}

Shortly after Einstein constructed general relativity (GR) in 1915, Weyl has advocated a generalization in that the very 
notion of length becomes path-dependent \cite{Weyl, Scholz}. In Weyl's theory, even if the lightcones retain the fundamental 
role as in GR, there is no absolute meaning of scales for space-time, so the metric is defined only up to proportionality. 
It is this property that we have a scale symmetry prohibiting the appearance of any dimensionful parameters and coupling constants 
in the Weyl theory. The main complaint against the Weyl's idea is that it inevitably leads to the so-called "second clock 
effect": The rate where any clock measures would depend on its history \cite{Penrose}. Since the second clock effect has 
not been observed by experiments, the Weyl theory might make no sense as a classical theory. 

However, viewed as a quantum field theory, the Weyl theory is a physically consistent and interesting theory and provides us with
a natural playground for constructing conformally invariant quantum field theories as shown in this article.\footnote{We
have already contructed the other scale invariant gravitational models \cite{Oda2, Oda3, Oda4}.}
Requiring the invariance under Weyl transformation is so strong that only quadratic curvature terms are allowed to
exist in a classical action, which should be contrasted with the situation of GR where any number of curvature terms could be 
in principle the candidate of a classical action only if we require the action to be invariant under diffeomorphisms.

Of course, we have a serious problem to be solved in future; the problem of unitarity. The lack of perturbative
unitarity is a common problem in the higher derivative gravity like the Weyl theory. However, it is expected that 
the Weyl gravity is asymptotically free, and the issue of the perturbative unitarity is closely relevant to infrared dynamics 
of asymptotic fields, so this problem becomes quite nontrivial. Provided that we can confine the ghosts in the Weyl theory
like in QCD, we would be free of the perturbative unitarity.

\begin{flushleft}
{\bf Acknowledgements}
\end{flushleft}
After submitting the paper \cite{Oda1}, we were informed by F. F. Faria of the reference \cite{Matsuo} 
where the effective potential of the conformal gravity in the Riemann geometry has been evaluated
by using the DeWitt-Schwinger method. We are grateful to F. F. Faria for this useful information.  
The present work was supported by JSPS KAKENHI Grant Number 16K05327.

\end{document}